\documentclass[12pt]{article}
\pdfoutput=1
\usepackage{draft}
\usepackage{comment}
\usepackage[normalem]{ulem}
\usepackage{hyperref}
\hypersetup{
 colorlinks=true,
 urlcolor=blue,
 anchorcolor=blue,
 citecolor=blue,
 filecolor=blue,
 linkcolor=blue,
 menucolor=blue,
 pagecolor=blue,
 linktocpage=true,
}
\usepackage{graphicx,color,subfig}
\usepackage{cite}
\usepackage{empheq} 
\usepackage[font=small,labelfont=bf]{caption} 

\usepackage[usenames,dvipsnames,table]{xcolor}
\graphicspath{{./figures/}}
\usepackage{tikz}
\usepackage{empheq}

\newcommand{\CE}{\mathcal{E}}

\newcommand{\CC}{\mathcal{C}}

\newcommand{\norm}[1]{\left\lVert#1\right\rVert}
\newcommand{\<}{\langle}
\renewcommand{\>}{\rangle}
\newcommand{\CO}{{\cal O}}

\usepackage{scalerel}
\usepackage{stackengine,wasysym}

\DeclareFontFamily{OT1}{pzc}{}
\DeclareFontShape{OT1}{pzc}{m}{it}{<-> s * [1.10] pzcmi7t}{}
\DeclareMathAlphabet{\mathpzc}{OT1}{pzc}{m}{it}

\definecolor{vert}{rgb}{0.1367 0.543 0.1367}

\def\({\left(}
\def\){\right)}
\newcommand{\nn}{\nonumber}

\begin{document}

\unitlength = .8mm

\begin{titlepage}

	  \begin{flushright}
	 \end{flushright} 

\begin{center}

 \hfill \\
 \hfill \\

\title{Higher $d$ Eisenstein Series and a Duality-Invariant Distance Measure}

~\vskip 0.01 in

\author{
	Nathan Benjamin$^{a}$, A. Liam Fitzpatrick$^{b}$
}
%
${}^a$\emph{\small Walter Burke Institute for Theoretical Physics, Caltech, Pasadena, CA 91125, USA}
\\
${}^b$\emph{\small Department of Physics, Boston University, Boston, MA 02215, USA}
~\vskip .2 in

\end{center}

\abstract{
The Petersson inner product  is a natural inner product on the space of modular invariant functions.  
We derive a formula, written as a convergent sum over elementary functions, for the inner product $E_s(G,B)$ of the real analytic Eisenstein series $E_s(\tau, \bar{\tau})$ and a general point in Narain moduli space.  We also discuss the utility of the Petersson inner product as a distance measure on the space of 2d CFTs, and apply our procedure to evaluate this distance in various examples.
}

\vfill

\end{titlepage}

\eject

\begingroup

\baselineskip .168 in
\tableofcontents

\endgroup

\section{Introduction and summary}
\label{sec:intro} 

The torus partition function $Z$ of a 2d CFT elegantly packages the infinite set of scaling dimensions of the theory into a single function which enjoys useful mathematical properties, chiefly modular invariance.  Modular invariance means that such partition functions can be thought of as functions on the fundamental domain, and the space of such functions has a natural inner product known as the Petersson inner product.  One can then attempt to study the space of 2d CFTs by studying the Petersson inner product of CFT torus partition functions with each other, or with other classes of functions.  A particularly natural set of functions to use is the eigenfunctions of the Laplacian on the upper-half plane, and the inner product of a function with this set is known as its Rankin-Selberg decomposition. Accurately evaluating the Petersson inner product in practice is not always straightforward, however.  In principle, it can be done numerically by brute force integration over the fundamental domain, but in practice this approach often suffers from numeric instabilities due to delicate cancellations between contributions from different integration regions. In this paper, we will apply a more efficient method from \cite{dixon1991moduli} for evaluating this inner product in the case where at least one of the functions in the inner product is the partition function of a Narain CFT.  

Our first application of this method will be to provide an algebraic expression, written as a sum over elementary functions, for the Rankin-Selberg decomposition of any Narain CFT.  More precisely, let $(f,g)$ be the Petersson inner product between functions $f$ and $g$. Let $Z(x,y)$ be the partition function of a Narain CFT with central charge $c$ on a torus with modulus $\tau = x + i y$, and define its reduced partition function $\hat{Z}(x,y) \equiv y^{c/2} |\eta(x,y)|^{2c} Z(x,y)$ where $\eta$ is the Dedekind eta function.  Finally, let $E_s(x,y)$ and $\nu_n(x,y)$, respectively, be the continuous and discrete eigenfunctions of the Laplacian on the upper-half plane; the former are known as the real analytic Eisenstein series, and the latter are known as the Maass cusp forms.  Then, we will derive an expression for $(\hat{Z}, E_s)$ and $(\hat{Z}, \nu_n)$.  Expressions for these inner products in the case of $c=1,2$ were given in \cite{Angelantonj:2011br,Obers:1999es,Obers:1999um}, and more recently in \cite{Benjamin:2021ygh} for $c=3,4$ in the decompactification limit when one of the target-space cycles is taken to be small.  The residue of the pole of this inner product at $s=1$ is a one-loop string amplitude, which was evaluated at $c=2$ in \cite{dixon1991moduli} and general $c$ in \cite{Kiritsis:1997em}.  We also provide explicit Mathematica code for the evaluation of $(\hat{Z}, E_s)$ with the arXiv preprint of this paper.

Our second application will be to compute the inner product between two different Narain CFT partition functions. In particular, we will be interested in the Petersson norm of their {\it difference},  $\Vert \hat{Z}_1 - \hat{Z}_2\Vert^2 \equiv (\hat{Z}_1 - \hat{Z}_2, \hat{Z}_1 - \hat{Z}_2)$.  Our main interest in this quantity is as a measure of the distance between the two CFTs.  Probably the most commonly used distance measure between two CFTs $\CC_1$ and $\CC_2$ is the Zamolodchikov metric, $d_Z(\CC_1, \CC_2)$.  The Zamolodchikov metric is appealing because it is fairly natural and can often be computed explicitly.  However, it also has a number of disadvantages.  First, it requires that the two CFTs $\CC_1$ and $\CC_2$ be connected by a moduli space parameterized by an exactly marginal deformation, since it is defined in terms of the correlators of that deformation.  So for most pairs of CFTs, their Zamolodchikov distance is simply undefined. Second, the Zamolodchikov metric between two dual CFTs does not typically vanish.  Since dual CFTs are, in a sense, the same theory, it seems more natural to define the distance between them to be zero.  By contrast, the ``Petersson distance'' $d_P(\CC_1, \CC_2)$ given by the Petersson norm of $\hat{Z}_1 - \hat{Z}_2$ does not suffer from either of these two drawbacks.  For instance, Narain CFTs with $c=1$ are parameterized by a single radius $r$, and we find that
\begin{equation}
d_P^2(r_1, r_2) \equiv \Vert \hat{Z}(r_1) - \hat{Z}(r_2) \Vert^2 = 4 \log \left( \frac{  \eta(i \frac{r_2}{r_1}) \eta(i \frac{r_1}{r_2}) \eta^2(i r_1 r_2)}{ \eta^2(i)\eta(i r_1^2)\eta(i r_2^2)} \right).
\label{eq:C1DP}
\end{equation}
This distance is invariant under duality $r \rightarrow 1/r$ of either $r_1$ or $r_2$, as can be seen using the identity $\eta(-1/\tau) = \sqrt{-i \tau} \eta(\tau)$, and vanishes when $r_1=r_2$.

Because $d_P(\CC_1, \CC_2)$ depends on the full partition functions of the two CFTs, it is unlikely that it can be calculated exactly except for solvable theories.  However, contributions from high dimension states should be small due to their Boltzmann suppression factor in the partition function, and so knowledge of the low-lying states should be sufficient to calculate $d_P$ between two CFTs to good approximation.  We will attempt to quantify more precisely the error from including in $d_P$ only the states below a threshold dimension $\Delta_*$ by using the asymptotic density of states imposed by modular invariance.

\section{Eisenstein Series in General $d$}
\label{app:Terras}

\subsection{Expression as Sum over Bessel Functions}

 Given two functions $f$ and $g$ defined on the fundamental domain $\mathcal{F} = SL(2,\mathbb{Z})\setminus H$,\footnote{As is standard, the upper half-plane $H$ can be parameterized by a complex coordinate $\tau=x +i y$ with $y>0$, and elements $\left\{ \left( \begin{array}{cc} a & b \\ c & d\end{array}\right) \Big| a,b,c,d \in \mathbb{Z}, ad-bc=1\right\}$ of $SL(2,\mathbb{Z})$ act on $\tau$ according to $\tau \rightarrow \frac{a \tau + b}{c \tau+d}$. } define the Petersson inner product to be 
\begin{equation}
(f,g) \equiv \int_{SL(2,\mathbb{Z})\setminus H} \frac{dx dy}{y^2} f(x,y) g^*(x,y).
\label{eq:Petersson}
\end{equation}
  As we review in appendix \ref{app:orbits}, the Petersson inner product of a Narain CFT partition function and another modular invariant function can be decomposed into sums over orbits of $SL(2, \mathbb{Z})$, so that for each orbit (aside from the trivial one) the integration region becomes either the upper half plane or the strip, and consequently the integrals simplify significantly and can often be done in closed form. The basic idea is to start with the reduced partition function written as the following manifestly modular invariant sum over lattice coordinates $(m^i,n^i) \in \mathbb{Z}^{2c}$:
\begin{equation}
\hat{Z}^{(c)}(x,y) = \sqrt{\det(G)} \sum_{m,n \in \mathbb{Z}} \exp\left( - \pi \frac{(m^i + n^i \tau) G_{ij} (m^j + n^j \bar{\tau})}{y} + 2 \pi i B_{ij}m^i n^j \right).
\end{equation}
An $SL(2,\mathbb{Z})$ transformation on $\tau$ can be compensated for by a rearrangement of the lattice coordinates, and consequently the lattice coordinates inherit an action of $SL(2,\mathbb{Z})$ acting on them, which defines the orbits. The reason the integration region becomes either the upper half plane or the strip is that for each orbit $\psi$, its stabilizer $\Gamma_\psi$ is either $\Gamma_\infty$ (the group generated by $T$ transformations in $SL(2,\mathbb{Z})$) or else trivial. Since the fundamental domain is the upper half-plane mod $SL(2,\mathbb{Z})$, the sum over each term within an orbit, integrated over the fundamental domain, can be `unfolded' into a single integral over a single representative of that orbit integrated over the upper-half plane mod the stabilizer of the orbit, which gives an integral over the strip or the upper half-plane for $\Gamma_\psi = \Gamma_\infty$ or $\Gamma_\psi = 1$, respectively.\footnote{A similar approach was taken in \cite{Bossard:2016hgy}, where the sum over lattice coordinates was decomposed into sums over orbits of $\Gamma_\infty$ to obtain a convergent sum for $(\hat{Z}^c, E_{1-s}$); see their (A.6). The reason  \cite{Bossard:2016hgy} used orbits of $\Gamma_\infty$, instead of the full $SL(2,\mathbb{Z})$ orbits, is that it allowed them to obtain a formula that reduces to a finite number of terms in the decompactification limit when the radius of a target space cycle is taken large.  }  For each orbit, we will denote the representative lattice coordinate for the orbit by $(m_\psi^i, n_\psi^i)$.

As our first application, we will compute the general $d$ Eisenstein series as a convergent sum.  That is, we will compute
\begin{equation}
(\hat{Z}^{(c)}, E_{1-s}),
\end{equation}
where $E_s$ is the real analytic Eisenstein series, and $\hat{Z}^{(c)} = y^{c/2} \eta^{c} Z$ is the reduced partition function of a central charge $c$ Narain CFT.  The main feature of the Eisenstein series $E_s$ that we will need is its Fourier expansion:
\begin{equation}
E_s(x,y) = y^s + \frac{\Lambda(1-s)}{\Lambda(s)} y^{1-s} + \sum_{j=1}^\infty 4 \cos(2 \pi j x) \frac{\sigma_{2s-1}(j)}{j^{s-\frac{1}{2}} \Lambda(s)} \sqrt{y} K_{s-\frac{1}{2}} (2\pi j y).
\end{equation}
We will substitute this expansion into the expression (\ref{eq:ReducedNarainInnerProduct}) for its inner product with the reduced Narain partition function.  First, consider the case where $\Gamma_\psi$ is just the identity element.  Then, the integrals can be performed in closed form, with the help of the following identities:
\begin{equation}
\begin{aligned}
\int_{-\infty}^\infty dx e^{ - \pi \frac{(m_\psi^i + n_\psi^i (x+i y)) g_{ij} (m_\psi^j + n_\psi^j (x-iy))}{y} } \cos(2\pi i j x) 
& = \sqrt{\frac{y }{g_{nn}} } e^{ - \frac{\pi}{g_{nn}} \Big( \frac{(g_{mm} g_{nn} - g_{nm}^2)}{y} + (j^2 + g_{nn}^2) y\Big)} \cos(2 \pi j  \frac{g_{mn}}{g_{nn}}) \\
\int_0^\infty \frac{dy}{y} e^{- \pi \left( \frac{a}{y} + y b \right)} y^{s} & = 2 (\sqrt{a/b})^{s} K_{s}(2\pi \sqrt{a b}),
\label{eq:UnfoldingDetNonzeroScalarPart} \\
\int_0^\infty \frac{dy}{y} e^{- \pi \left( \frac{a}{y} + y (b + \frac{j^2}{b}) \right)} K_{s}(2 \pi j y) &= 2 K_{s}(2 \pi \sqrt{ab}) K_{s}(2 \pi j \sqrt{a/b}).
\end{aligned}
\end{equation}
where we have introduced
\begin{equation}
g_{nn} \equiv n_\psi^i G_{ij} n_\psi^j, \qquad g_{nm} \equiv n_\psi^i G_{ij} m_\psi^j , \qquad g_{mm} \equiv m_\psi^i G_{ij} m_\psi^j, \qquad b_{mn} \equiv m_\psi^i B_{ij} n_\psi^j.
\end{equation}
and we will take $a=\frac{g_{mm} g_{nn} - g_{mn}^2}{g_{nn}}$ and $b=g_{nn}$.

Putting it all together, when $\Gamma_\psi$ is trivial, we have
\begin{equation}
\begin{aligned}
( \hat{Z}^c_\psi, E_{1-s}) &= \sqrt{\frac{\det g}{g_{nn}}} 2 K_{s-\frac{1}{2} }(2 \pi \sqrt{a b}) e^{2\pi i b_{mn}} \Big( (\sqrt{a/b})^{s-\frac{1}{2}} + \frac{\Lambda(s-\frac{1}{2})}{\Lambda(\frac{1}{2}-s)} (\sqrt{a/b})^{\frac{1}{2}-s} \\
& + \sum_{j=1}^\infty 4 \cos( 2 \pi j \frac{g_{mn}}{g_{nn}}) K_{s-\frac{1}{2}}(2 \pi j \sqrt{a/b}) \frac{\sigma_{2s-1}(j)}{j^{s-\frac{1}{2}} \Lambda(\frac{1}{2}-s)}
\Big) .
\label{eq:TrivStabOverlap}
\end{aligned}
\end{equation}

The other two possibilities for $\Gamma_\psi$ are that $\Gamma_\psi = \Gamma_\infty$ and $\Gamma_\psi = SL(2, \mathbb{Z})$.  As discussed around (\ref{eq:AllDsVanish}), if $\Gamma_\psi = \Gamma_\infty$, then we can choose the representative $(n_\psi^i, m_\psi^i)$s so that $n_\psi^i=0$. Therefore, in this case,
only the $j=0$ term of the Eisenstein series contributes, and moreover all dependence on $B$ drops out:
\begin{equation}
\begin{aligned}
\sum_{ \{ \textrm{orbits } \psi \ | \ \Gamma_\psi = \Gamma_\infty \}} (\hat{Z}_\psi^{c}, E_{1-s}) & = 2 \sqrt{\det(G)} \left( E_{c}(G,1-s) + \frac{\Lambda(1-s)}{\Lambda(s)} E_c(g,s) \right),  \\
E_c(G,s) \equiv & \frac{1}{2} \int_0^\infty \frac{dy}{y} y^s \sum_{m \in \mathbb{Z}^n}' e^{- \pi y m^i G_{ij} m^j} .
\end{aligned}
\end{equation}
This term is exactly the one worked out in \cite{terras1973bessel},
as we briefly summarize in Appendix \ref{app:Terras}. The final result in \cite{terras1973bessel} is the following recursive formula for $E_c(G,s)$:
\begin{equation}
\begin{aligned}
E_{n_1+n_2}(G,s)  =& E_{n_2}(G_2, s) + \frac{1}{\sqrt{\det G_2}} E_{n_1}(T_1, s - \frac{n_2}{2}) \\
        & +\frac{1}{\sqrt{\det G_2}}  \sum_{a_1 \in \mathbb{Z}^{n_1} - 0 \atop b_2 \in \mathbb{Z}^{n_2}-0 } \left( \frac{G_2^{-1}[b_2]}{T_1[a_1]} \right)^{\frac{2s-n_2}{4}}  e^{- 2 \pi i b_2 Q a_1} K_{s - \frac{n_2}{2}} (2\pi \sqrt{ T_1[a_1] G_2^{-1}[b_2]}) . 
        \label{eq:TerrasFormula}
\end{aligned}
\end{equation}
Finally, the only case with $\Gamma_\psi=SL(2,\mathbb{Z})$ is the trivial orbit, with $n_\psi^i = m_\psi^i=0$, which is discarded according to the prescription in \cite{zbMATH03796039}. To get the full $c$-dimensional Narain Eisenstein series, we sum over all orbits:
\begin{equation}
\boxed{(\hat{Z}^c, E_{1-s}) = 2 \sqrt{\det(G)} \left( E_{c}(G,1-s) + \frac{\Lambda(1-s)}{\Lambda(s)} E_c(G,s) \right)+ \sum_{\{ \textrm{orbits } \psi \  |  \ \Gamma_\psi=1 \}} (\hat{Z}_\psi^c, E_{1-s}),}
\label{eq:EisFinal}
\end{equation}
with $(\hat{Z}_\psi^c, E_{1-s})$ in the last term given in (\ref{eq:TrivStabOverlap}). 
In appendix \ref{app:orbits}, we discuss in  detail how to efficiently perform a numeric sum over orbits in the last term.   More precisely, if the full set of lattice coordinates is written as
\begin{equation}
N_{ij} = \left( \begin{array}{cccc} n^1 & n^2 & \dots & n^c \\ m^1 & m^2 & \dots & m^c \end{array} \right),
\end{equation}
then every determinant $D_{ij}$ of every $2\times 2$ submatrix is an invariant under $SL(2,\mathbb{Z})$, and we can organize the sum over orbits as a sum over these invariants. For a given $c$, there are $\frac{c(c-1)}{2}$ different determinants $D_{ij}$, and the algorithm outlined in the appendix first sums over the values of these determinants.  For any given choice of all the determinants, there will always be a finite number of possible orbits consistent with those values of $D_{ij}$; in fact, for most choices of all the $D_{ij}$s there will not be any orbit consistent with all of them.  The algorithm we describe takes any choice of all the determinants and  returns a list with one representative set of lattice coordinates for every consistent orbit; whenever they are not consistent, this list is simply empty. 



\begin{figure}[th!]
\begin{center}
\includegraphics[width=0.49\textwidth]{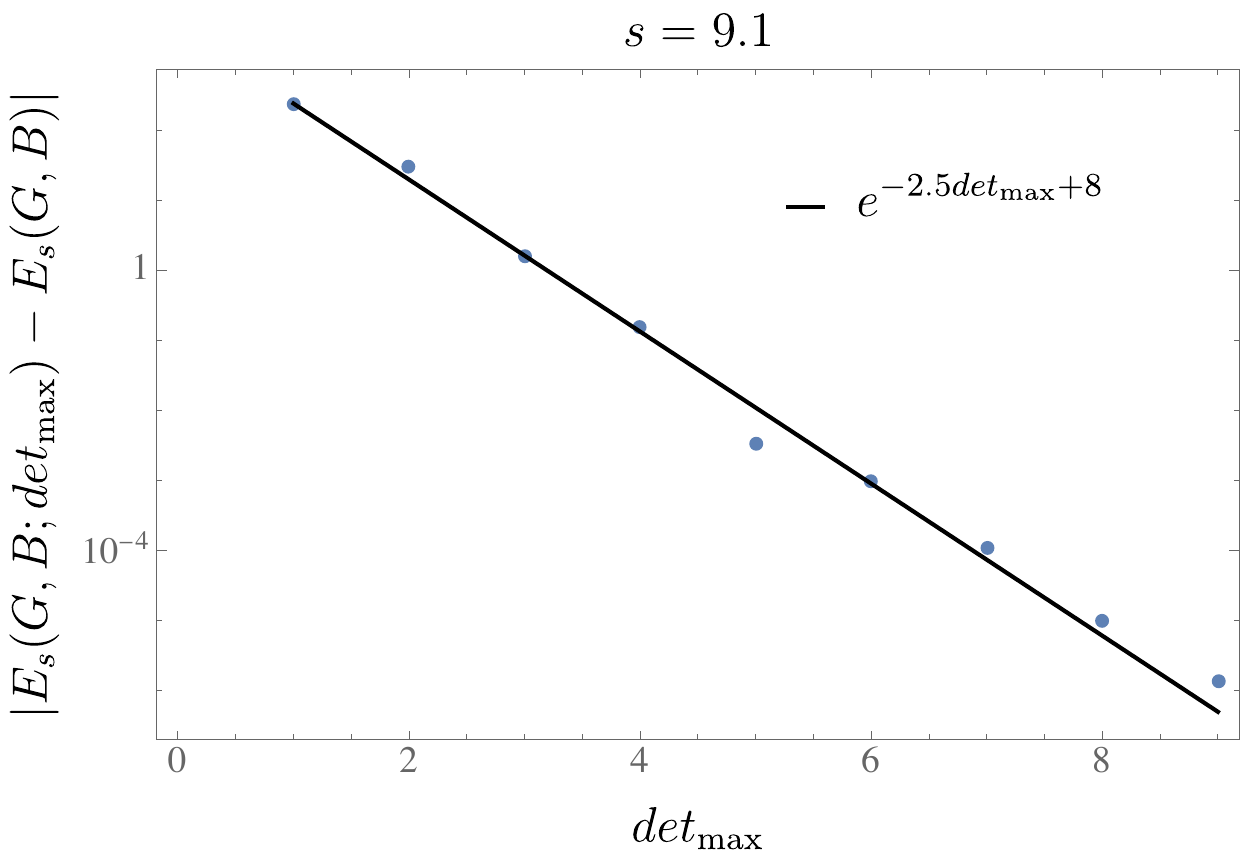}
\includegraphics[width=0.49\textwidth]{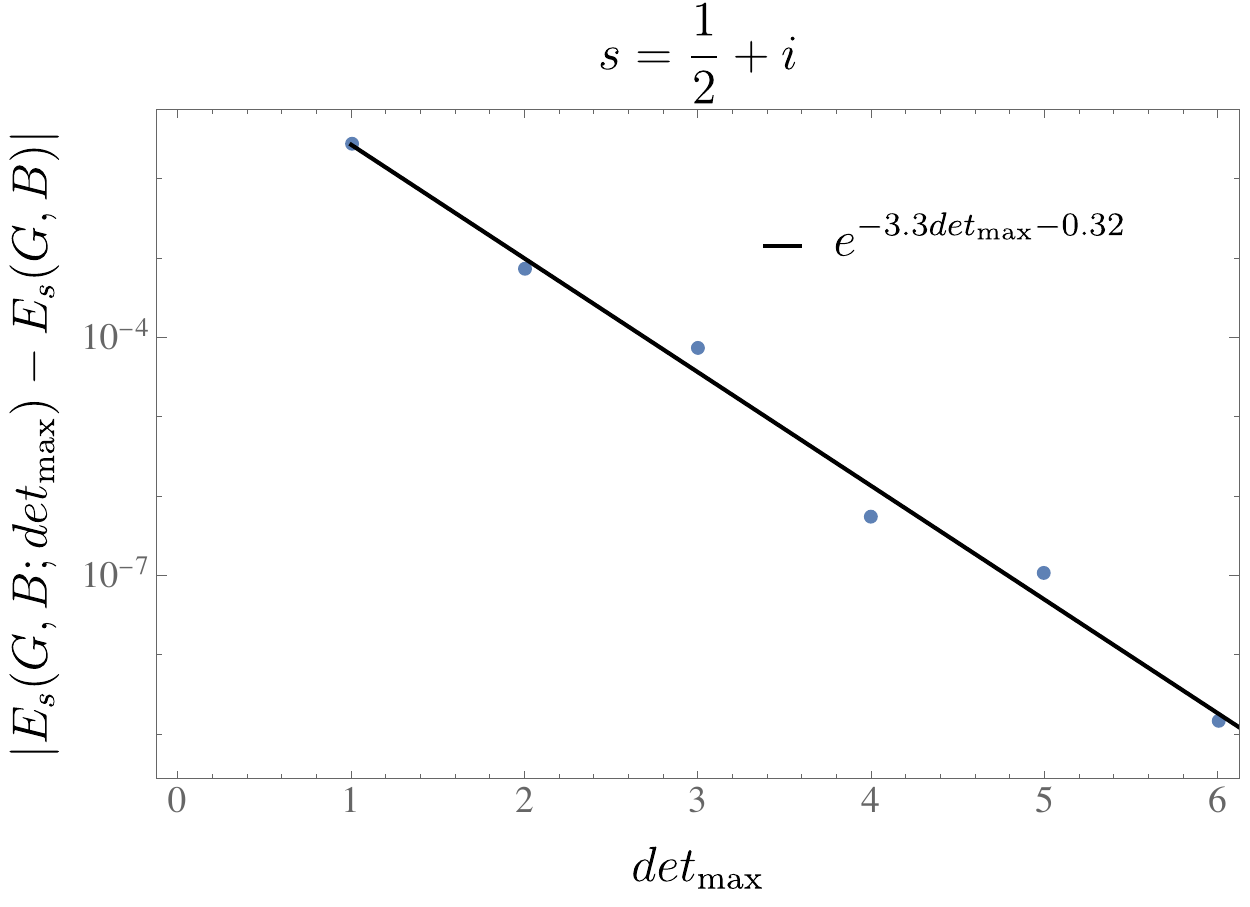}
\caption{Convergence with the cutoff $det_{\rm max}$ of the $c=3$ Narain CFT Eisenstein series at the $SU(4)_1$ point, at two values of $s$.  }
\label{fig:convergence}
\end{center}
\end{figure}

The sum over orbits converges by virtue of the exponential decay of the Bessel function $K_\nu(z)$ at large $z$. 
 To evaluate the sums in practice on a computer, we truncate the sum to only include orbits with $D_{ij}$ less than some maximum value $det_{\rm max}$:
\begin{equation}
| D_{ij}| \le det_{\rm max}.
\end{equation}
We therefore expect that as we increase the cutoff $det_{\rm max}$, the sum should converge exponentially, with an error $\sim e^{- \# det_{\rm max}}$ for some value of $\#$. In Fig.~\ref{fig:convergence}, we exhibit this numerically for $c=3$ at the point in moduli space with the largest gap in the spectrum of scalar operators, 
\begin{equation}
G = \left( \begin{array}{ccc} \frac{1}{2} &  0 & 0\\ 0 & \frac{1}{2} & 0\\ 0& 0& \frac{1}{2} \end{array} \right), \qquad B = \left( \begin{array}{ccc}  0 & -\frac{1}{2} & - \frac{1}{2}  \\  \frac{1}{2} & 0 & - \frac{1}{2}  \\ \frac{1}{2}  & \frac{1}{2}  & 0 \end{array} \right),
\end{equation}
 which turns out to be the $SU(4)_1$ WZW model.

 In the arXiv submission for this paper, we include Mathematica code that evaluates (\ref{eq:EisFinal}) in the case of $c=3$ for an arbitrary $G$ and $B$.

\section{Distance Measure}

\subsection{Motivation}

The idea of a `space of CFTs' can be a useful concept: some statements about dynamics of a theory are naturally stated as an average over theories, and sometimes a specific theory lies at the boundary of a space of all possible theories in a way that can be used to compute observables in that theory. Having a metric on such a space can help clarify its structure.  Probably the standard metric of choice for the space of CFTs is the Zamolodchikov metric, which is defined on CFTs connected to each other by exactly marginal deformations in terms of the two-point function of those deformations. That is, given a CFT with a set $\{ \CO_i \}$ of exactly marginal deformations, we can perturb it by adding $\lambda^i \int d^d x \CO_i$ to its action.  The local metric on this space of couplings is  
\begin{equation}
g_{i j} d\lambda^i d \lambda^j \equiv x^{2d} \< \CO_i(x) \CO_j(0)\> d \lambda^i d \lambda^j.
\end{equation}
For instance, the $c=1$ theory of a single compact free scalar field $\varphi$ of radius $r$ ($\varphi \cong \varphi+2\pi r$) has a single exactly relevant deformation $\CO \equiv (\partial \varphi)^2$.  Adding $\lambda \int d^2 x \CO$ to the action  is equivalent to changing the compactification radius $r$ to $r' = \sqrt{1+\lambda}r \approx r + r \frac{\lambda}{2} + \CO(\lambda^2)$, since it can be  absorbed into a rescaling of $\phi$ to $\phi ' = \sqrt{1+\lambda} \phi$.  The two-point function of $\CO$ in a canonical normalization is just $\<\CO(x) \CO(0)\> = x^{-4}$.  Therefore, the Zamolodchikov metric in this case is
\begin{equation}
g_{11} d\lambda^2 = d\lambda^2 = \frac{dr^2}{r^2}.
\end{equation}

The Zamolodchikov metric has many appealing features, and is a natural metric if we think about the space of CFTs in terms of exactly marginal deformations.\footnote{See also \cite{Stout:2022phm} for an information-theoretic perspective on the Zamolodchikov metric and its relation to distance conjectures.}  However, there are reasons it might still be useful to have other metrics at our disposal.  One  disadvantage of the Zamolodchikov metric is that the distance between dual theories does not vanish.  The $c=1$ case above is dual under $r \leftrightarrow r^{-1}$, so two theories related by this duality are in a sense the same theory. If we are trying to use a metric to help characterize the space of CFTs, a fairly basic property we might demand of such a metric is that the distance of a theory to itself should be zero.  One can try to engineer this condition in the case of the Zamolodchikov metric by identifying theories that are dual to each other, but this does not resolve the issue in the vicinity of a duality fixed point, where for instance  the infinitesimal Zamolodchikov distance between $r=1+d\delta$ and its dual description at $r^{-1}=1-d \delta$ is still $d\delta>0$; in geometric language, the orbifold creates an orbifold singularity at the fixed point.

Perhaps a more practical disadvantage of the Zamolodchikov metric is that most CFTs are not connected by exactly marginal deformations. Indeed, exactly marginal deformations rarely if ever show up at all outside of supersymmetric  and two-dimensional cases.  
 Instead, if we want to work with a CFT metric that can apply to all CFTs, it is more natural to work with the `CFT data', i.e., the dimensions of operators and their OPE coefficients.  A more general approach based on CFT data would also have the advantage that dual theories are automatically seen as being equivalent, since their CFT data by definition is the same.   Various proposals for a CFT metric based purely on CFT data were proposed in \cite{Douglas:2010ic} with this motivation in mind, which were some of the inspiration for this work.

 In this section, we will define a new metric between 2d CFTs $\CC_1$ and $\CC_2$, based on the Petersson norm of the difference between their (reduced) torus partition functions:
 \begin{equation}
 d_P^2(\CC_1, \CC_2 ) \equiv \norm{ \hat{Z}_1 - \hat{Z}_2}^2.
 \label{eq:Petersson2}
 \end{equation}
This definition of the distance manifestly vanishes between two self-dual theories, since by definition they have the same spectrum of dimensions and spins and therefore the same partition function.  Moreover, it does not depend on the existence of a moduli space, and can be evaluated between any two 2d CFTs.  It also seems fairly natural -- the torus partition function is probably the most natural way to encapsulate the information of the spectrum of a theory, and in 2d CFTs the torus partition function is a function of the fundamental domain, for which the natural norm is the Petersson norm.  

We foresee two objections one might nevertheless make to the definition (\ref{eq:Petersson2}).  The first is that, although it vanishes between  two equivalent theories, the converse is not true: there are known 2d CFTs with the same spectrum but different OPE coefficients.\footnote{For example $E_8 \times E_8$ and $SO(32)$ lattices are famously isospectral. Moreover, for holomorphic CFTs with $c_L=0, c_R=24$, there are several known examples of different CFTs with the same torus partition function \cite{Schellekens:1992db}.} This is probably not a very serious drawback since for a given central charge, the size of such families of inequivalent CFTs with the same spectrum is usually small compared to the number of all CFTs at that central charge, and it is still fruitful to first characterize the space of partition functions.\footnote{Moreover, one could view (\ref{eq:Petersson2}) as the simplest in a family of metrics, where instead of using the torus partition function one uses the partition function on higher genus surfaces.  As the genus of the surface increases, the partition function becomes increasingly sensitive to CFT data. } A more serious objection is that it seems like  (\ref{eq:Petersson2}) will be impossible to evaluate in practice outside of very special cases where $\CC_1$ and $\CC_2$ are solvable so that one can obtain their full spectrum.  However, if we are content to compute (\ref{eq:Petersson2}) to high precision, rather than computing it exactly, then we do not need to know the full spectrum.  The key point is that, because of Boltzmann suppression,\footnote{The highest temperature in the fundamental domain is at the cusps, where $\textrm{Im}(\tau)= \frac{\sqrt{3}}{2}$. } the contribution from the lightest states should dominate the partition function.  

\begin{figure}
  \centering
      \includegraphics[width=0.5\textwidth]{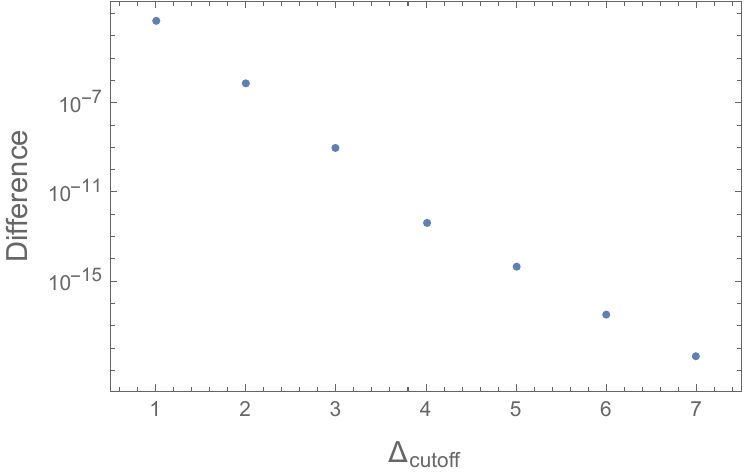}
  \caption{Plot of the error in the distance of $r=1$ and $r=2$ Narain theories as a function of $\Delta_{\text{cutoff}}$.}
  \label{fig:z1z2}
\end{figure}

Consequently, in practice, if we only know the CFT spectrum up to some large but finite energy $\Delta \gg c$, we can compute the Petersson distance to exponentially good accuracy in $\Delta$. More precisely, states with energy $\Delta$ have a Boltzmann suppression of $e^{-2\pi \Delta y}$ when integrated over the fundamental domain. We can bound the integral of a single state by its value at the minimum value of $y$ in $\mathcal{F}$ (namely, $\frac{\sqrt{3}}2$) times the volume of $\mathcal{F}$. This shows gives an exponential falloff in $\Delta$. Since the number of states at large $\Delta$ grows roughly as $e^{2\pi\sqrt{\frac{\Delta c}3}}$, this doesn't change the exponential falloff in $\Delta$. See Fig. \ref{fig:z1z2}.

\subsection{Application to Narain CFT Moduli Space}

We describe two different strategies one can use to compute the distance between two Narain CFT partition functions. The first strategy is to first unfold one of the two Narain partition functions in the inner product $(\hat{Z}_1, \hat{Z}_2)$.  After unfolding, say, $\hat{Z}_1$, we will be able to perform all integrations in closed form, except for the contribution from the trivial orbit of $\hat{Z}_1$ where the `unfolded' integration region is still $\mathbb{F}$; for this term only, we can also unfold $\hat{Z}_2$.   The second strategy is we use the fact that the product of two Narain CFT partition functions at central charge $c$ is another Narain CFT partition function at central charge $2c$.  The advantage of the second strategy is that it reduces to the special case of the inner products considered in the previous section, of a Narain CFT partition function with an Eisenstein function $E_s$, where we simply take the contribution from the pole at $s=0$. 

\subsubsection{$c=1$ Example}

Consider the Petersson distance $d_P^2(r_1, r_2)$ between two $c=1$ Narain CFTs with radii $r_1, r_2$ respectively:
\begin{equation}
d_P^2 = \int_{\mathbb{F}} \frac{dx dy}{y^2} \left| \hat{Z}_1(x,y) - \hat{Z}_2(x,y)\right|^2, \quad \hat{Z}_i(x,y) = \sqrt{y} \sum_{m,n = -\infty}^\infty e^{ - \pi y \left( m^2 r_i^2 + \frac{n^2}{ r_i^{2}}\right) + 2 \pi i m n x}.
\label{eq:C1Distance}
\end{equation}
We will first evaluate this expression by applying equation (\ref{eq:C1Unfolded}) to the difference $(\hat{Z}_1, g)-(\hat{Z}_2, g)$ with $g=\hat{Z}_1- \hat{Z}_2$:
\begin{equation}
\begin{aligned}
d_P^2
&= (r_1 - r_2)\int_{\mathbb{F}} \frac{dx dy}{y^2} (\hat{Z}_1 - \hat{Z}_2) \\
&+ \int_{-\frac{1}{2}}^{\frac{1}{2}} dx \int_0^\infty \frac{dy}{y^2}\left[r_1 \left(\vartheta\left(\frac{r_1^2}{y}\right) -1\right)-r_2 \left(\vartheta\left(\frac{r_2^2}{y}\right) -1\right)\right] (\hat{Z}_1 - \hat{Z}_2) , 
\label{eq:C1Distance2}
\end{aligned}
\end{equation}
where $\vartheta(t)$ is the theta function:
\begin{equation}
\vartheta(t) \equiv \sum_{m=-\infty}^\infty e^{-\pi m^2 t},
\end{equation}
which satisfies $\vartheta(t) = t^{-\frac{1}{2}} \vartheta(t^{-1})$. 

The second term on the RHS of (\ref{eq:C1Distance2}) simplifies because the $dx$ integral projects onto the scalar sector. It can be evaluated term-by-term in the expansion of the $\vartheta$ functions:
\begin{equation}
\begin{aligned}
& \int_{-\frac{1}{2}}^{\frac{1}{2}} dx \int_0^\infty \frac{dy}{y^2}\left[r_1 \left(\vartheta\left(\frac{r_1^2}{y}\right) -1\right)-r_2 \left(\vartheta\left(\frac{r_2^2}{y}\right) -1\right)\right] (\hat{Z}_1 - \hat{Z}_2) \\
&= 4\sum_{m,n=1}^\infty \left( \frac{e^{-2 \pi  m n r_1^2}-e^{-\frac{2 \pi  m n r_1}{r_2}}-e^{-\frac{2 \pi  m n r_2}{r_1}}-2 e^{-2 \pi  m n r_1 r_2}+e^{-2 \pi  m n r_2^2}+2 e^{-2 \pi  m n}}{m} \right) \\
&= -\frac{\pi  \left(r_1-r_2\right){}^2 \left(r_1 r_2-1\right)}{3 r_1 r_2}+8 \log \left( \frac{ \sqrt{ \eta(i \frac{r_2}{r_1}) \eta(i \frac{r_1}{r_2})} \eta(i r_1 r_2)}{\eta(i) \sqrt{\eta(i r_1^2)\eta(i r_2^2)}} \right).
\end{aligned}
\end{equation}
 By contrast, the first term on the RHS of (\ref{eq:C1Distance2}) is still an integral over $\mathbb{F}$, so to simplify it we unfold  each integral over $\hat{Z}_i$ as follows:
\begin{equation}
 \int_{\mathbb{F}} \frac{dx dy}{y^2} \hat{Z}_i = r_i \int_{\mathbb{F}} \frac{dx dy}{y^2}+ \int_0^\infty \frac{dy}{y^2} r_i \left(\vartheta\left(\frac{r_i^2}{y}\right)-1\right)=  \frac{\pi}{3} (r_i + r_i^{-1}). 
\end{equation}
Combining both terms, we obtain the result (\ref{eq:C1DP}) as claimed in the introduction:
\begin{equation}
d_P^2(r_1, r_2) = 8 \log \left( \frac{ \sqrt{ \eta(i \frac{r_2}{r_1}) \eta(i \frac{r_1}{r_2})} \eta(i r_1 r_2)}{\eta(i) \sqrt{\eta(i r_1^2)\eta(i r_2^2)}} \right).
\label{eq:C1DP2}
\end{equation}

Alternatively, we can use the fact that the product of two $c=1$ Narain  CFT partition functions is the partition function of a $c=2$ partition function on the product locus. In \cite{dixon1991moduli} (see also \cite{Benjamin:2021ygh}), the following expression was obtained for the average over $\mathbb{F}$ of a $c=2$ partition function $\hat{Z}^{(c=2)}$:
\begin{equation}
\int_{\mathbb{F}} \frac{dx dy}{y^2} \hat{Z}^{(c=2)}  = \frac{\pi}{3} \left( \widehat{E}_1(\rho) + \widehat{E}_1(\sigma) + \delta\right), 
\end{equation}
where $\delta$ is a numeric  constant that will cancel in our applications, and 
\begin{equation}
\widehat{E}_{1}(\tau) = \lim_{s\rightarrow1}\left[ E_s(\tau) -\frac{3}{\pi(s-1)}\right] = y - \frac{3 \log y}{\pi} + \frac{12}{\pi} \sum_{m=1}^\infty \frac{\sigma_1(m) e^{-2 \pi m y} \cos(2 \pi m x) }{m}.
\end{equation}  
From (\ref{eq:C2Modulus}), one can read off that the product locus of a $c=2$ Narain CFT that factors into two $c=1$ Narain CFTs with radii $r_1$ and $r_2$ (i.e. $G_{11}=r_1^2, G_{22}=r_2^2, G_{12}=G_{21}=0$) is $\rho= i r_1 r_2, \sigma = i r_2/r_1$.  We therefore find
\begin{equation}
d_P^2(r_1, r_2) = \frac{\pi}{3} \left( \widehat{E}_1(ir_1^2)+\widehat{E}_1(ir_2^2) + 2 \widehat{E}_1(i) - 2 \widehat{E}_1(ir_1 r_2)- 2 \widehat{E}_1(ir_2/r_1)\right).
\label{eq:C1DP3}
\end{equation}
Using the identities $\sum_{m=1}^\infty \frac{\sigma_1(m)}{m} e^{-2 \pi y m} = -\frac{\pi y}{12}-\log \eta(i y)$ and $\eta(i r^{-1})/\eta(i r) = \sqrt{r}$, one can easily demonstrate that (\ref{eq:C1DP3}) is equal to  (\ref{eq:C1DP2}).

Sometimes, it is more natural to consider the infinitesimal local distance of a metric than its finite form.  The infinitesimal line element corresponding to $d_P^2(r_1, r_2)$ is
\begin{equation}
ds^2 = \left[ \frac{1}{2} \frac{d^2}{dr'^2} d_P^2(r, r')\right]_{r=r'} dr^2 = \frac{\pi^2}{3} \left( r^2 D_2 \widehat{\CE}_2(i r^2) - \frac{2 \eta^8(i)}{r^2} \right)dr^2,
\end{equation}
where $D_2\widehat{\CE}_2(\tau)= \left( \frac{i}{\pi} \partial_\tau + \frac{1}{\pi \tau_2} \right) \widehat{\CE}_2(\tau) = \frac{1}{6} \CE_4(\tau) - \frac{1}{6} \widehat{\CE}_2^2(\tau)$ is the modular covariant derivative acting on the modular form $\widehat{\CE}_2 = \CE_2 - \frac{3}{\pi \tau_2}$, and we have introduced the  notation $\CE_n(\tau)$ for the holomorphic Eisenstein series.\footnote{The holomorphic Eisenstein series are usually denoted by $E_n(\tau)$, but here we use $\CE_n(\tau)$ to distinguish them from the real analytic Eisenstein series.}  At large $r$, $D_2 \widehat{\CE}_2(ir^2)  \sim \frac{1}{\pi r^2}$, and
\begin{equation}
ds^2 \stackrel{r \gg1}{\sim} \frac{\pi}{3} dr^2.
\end{equation}

\begin{figure}
  \centering
      \includegraphics[width=0.49\textwidth]{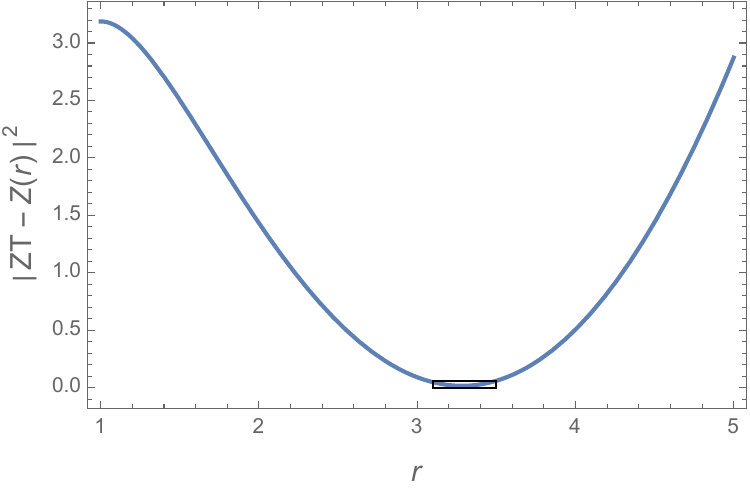}
      \includegraphics[width=0.49\textwidth]{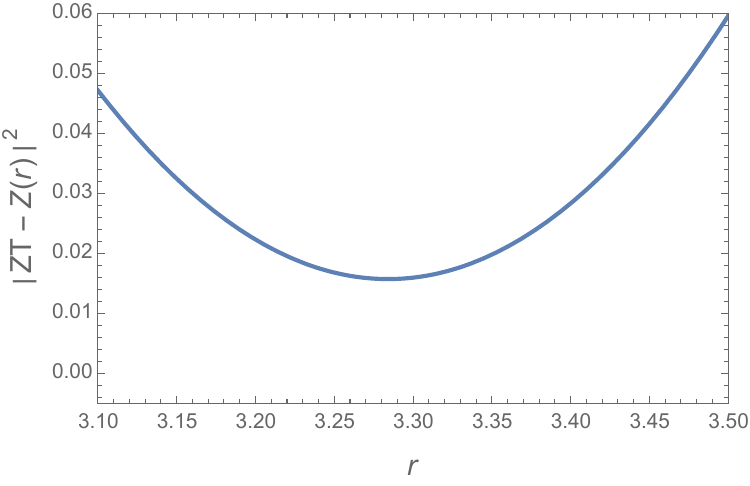}
  \caption{Distance squared between $c=1$ compact free bosons at radius $r$ to the exceptional $T$ $c=1$ CFT, as a function of radius. (Right image is zoomed-in version of left simply to show the distance never vanishes.)}
  \label{fig:c1tdist}
\end{figure}

Finally we note that this metric allows us to define the distance between CFTs not necessarily connected by marginal operators. For example, we can explicitly compute the distance from a single $c=1$ compact free scalar field to an exceptional $c=1$ CFT that has no marginal operators \cite{Ginsparg:1987eb}.\footnote{This question for instance was explicitly raised in \cite{Douglas:2010ic}.} There are believed to be three exceptional, isolated $c=1$ theories, commonly denoted $T$, $O$, and $I$. Since their partition functions are linear combinations of partition functions of $c=1$ compact free scalar fields, we can easily compute the distance. In Fig. \ref{fig:c1tdist}, we plot the distance from a $c=1$ free scalar field at radius $r$ to the $c=1$ $T$ exceptional CFT. 

\subsubsection{Decompactification Limit and Distance Conjectures}

Next, consider the inner product between two central charge $c$ Narain CFTs in the limit that one of the cycles is taken to be large for both of them.  As discussed above, the computation of their Petersson distance is equivalent to computing the average over $\mathbb{F}$ of a $c'=2c$ Narain CFT at a point in moduli space that is the product of two central charge $c$ points, in this case with a lattice that factorizes as $\Gamma_{c'} = (\Gamma_1 \times \Gamma_{c-1} ) \times (\Gamma_1 \times \Gamma_{c-1})$.  Let $R_1,R_2$ be the radius of the two large $S^1$s, and let $\hat{Z}^{(2c-2)}$ be the reduced partition function
of the central charge $2c-2$ Narain lattice which is the product of the two $\Gamma^{c-1}$ factors.  Then we can apply a decompactification formula from \cite{Benjamin:2021ygh} treating the two large radius components as a single $c=2$ Narain CFT with $\rho = i R_1 R_2$ and $\sigma = i R_1/R_2$, with the following result:
\begin{equation}
(\hat{Z}^{(2c)},1) = R_1 R_2 \left[ (\hat{Z}^{(2c-2)}, 1) + \left\{ \begin{array}{cc}   (R_1 R_2)^{c-2} \Lambda(c-1)E_{c-1}(i R_1/R_2) & c >2 \\ 
\frac{\pi}{3} \widehat{E}_1(i R_1/R_2) + \log(R_1 R_2)-3.77 & c=2 \end{array} \right\} \right]+\dots ,
\end{equation}
where $\dots$ indicates corrections that are nonperturbatively small in $R_1 R_2$.

In the special case of $c=2$, the factor $(\hat{Z}^{(2c-2)},1)$ is known in closed form:
\begin{equation}
(\hat{Z}^{(2)}_1, \hat{Z}^{(2)}_2) = (\hat{Z}^{(4)}, 1) \approx R_1 R_2 \left(\frac{\pi}{3} (\widehat E_1(i R_1/R_2)+\widehat E_1(i r_1 r_2)+\widehat E_1( ir_1/r_2))+\log\rho_2^{(1)} -6.63\right),
\end{equation}
In terms of these inner products, the Petersson distance is simply $d_P^2 \equiv \Vert \hat{Z}_1 - \hat{Z}_2 \Vert^2 = (\hat{Z}_1, \hat{Z}_1) - 2 (\hat{Z}_1, \hat{Z}_2)+(\hat{Z}_2, \hat{Z}_2)$.

An intriguing question about distances between CFTs is what types of behavior only arise at infinite distance.  In particular, in \cite{Baume:2020dqd, Perlmutter:2020buo} (partially inspired by the gravity distance conjecture from \cite{Ooguri:2006in}), the authors put forward a `CFT distance conjecture' that all infinite distance points have a tower of massless higher spin states. 
In terms of the Zamolodchikov metric, it was proven that all higher spin points are indeed at infinite distance in $d>2$ in \cite{Baume:2023msm}, though whether or not the converse is true remains an open question. 
In $d=2$, the status of the conjecture is less clear because the implications of an exactly conserved higher spin current differs greatly between $d=2$ and $d>2$, but one natural extension is that infinite distances are associated with massless scalars \cite{Kontsevich:2000yf, Acharya:2006zw}. 
  Here we will make some observations on its status in the context of the Petersson distance.

Because the Petersson distance is defined  in terms of the torus partition function, it is directly related to the spectrum of states, which makes it easier to see what effect the presence of light states might have on this distance.  Consider a family of CFTs parameterized by $\lambda$, and let us ask what happens if there exists a scalar state with dimension $\Delta(\lambda)$  such that $\Delta(\lambda)$ vanishes as $\lambda$ approaches a critical value which we can set to zero without loss of generality.   At large $y$, the leading and subleading terms of the reduced partition function are the contribution from the vacuum and the light state with dimension $\Delta(\lambda)$, respectively:
\begin{equation}
\hat{Z}(\lambda) \stackrel{y \sim \infty}{\sim} y^{\frac{c}{2}} + N y^{\frac{c}{2}} e^{- 2 \pi \Delta(\lambda) y} + \dots,
\end{equation}
where $N$ is the degeneracy of the light state.  A short calculation shows that the contribution from this state to the Petersson distance is
\begin{equation}
d_P^2(\hat{Z}(\lambda), \hat{Z}(\lambda+d \lambda)) \approx N^2 \frac{4 \pi^2 \Gamma(1+c)}{(4 \pi \Delta(\lambda))^{c+1}} d \Delta^2,
\end{equation}
where $d \Delta = \Delta(\lambda+ d \lambda) - \Delta(\lambda)$. By inspection, the integrated distance to a point with $\Delta(\lambda)=0$ diverges for $c\ge 1$. 

\section{Future Directions}

Our primary focus in this paper has been a technical one, of how to use the manifest modular invariance of the expression (\ref{eq:PoissResum}) for Narain CFT partition functions in order to simplify their Petersson inner products with other modular functions.  Here we make some comments about what the resulting expressions might be useful for.  Consider the expression (\ref{eq:EisFinal}) for the overlap between the partition function and the real analytic Eisenstein series.  One interesting property of this function is that its large $s$ limit is determined by the dimension $\Delta_{\rm sgap}$ of the lightest scalar state (after the vacuum), i.e., the `scalar gap' or, as it is commonly known, the `sgap' for short:
\begin{equation}
(\hat{Z}^c, E_{1-s}) \stackrel{s\sim \infty}{\sim} \frac{\Gamma(s+ \frac{c-1}{2})}{\pi^{s + \frac{c-1}{2}}} \frac{N_{\rm sgap}}{(2\Delta_{\rm sgap})^{s + \frac{c-1}{2}}} ,
\end{equation}
where $N_{\rm sgap}$ is the number of scalar states with dimension $\Delta_{\rm sgap}$.  So, the expression (\ref{eq:EisFinal}) might potentially be useful as a tool to learn about the sgap as a function of Narain moduli space, and where it is maximized.  For instance, for a fixed value of the metric $G$ field, consider the value of $B$ that maximizes $\Delta_{\rm sgap}$.  In this case, the first term on the RHS in (\ref{eq:EisFinal}) is constant, and variations in $\Delta_{\rm gap}$ as a function of $B$ only require evaluating the second term.  Moreover, when doing a numeric search for the extremum, one could likely use lower truncations in the sums (or perhaps smaller values of $s$) early in the search, improving speed at the cost of accuracy, and increase the truncation to improve the accuracy of the calculation as one gets closer to the extremum.  

Another interesting direction would be to understand the overlap of Narain CFT partition functions with Maass cusp forms, rather than with Eisenstein series.\footnote{Recall that the Maass cusp forms, denoted $\nu_n(\tau)$ with $n \in\mathbb N$, are a discrete set of eigenfunctions of the Laplacian on $\mathbb{H}/SL(2,\mathbb Z)$ with sporadic eigenvalues. The formulas written here are for even-parity Maass cusp forms, but essentially identical ones occur for the odd ones. See \cite{Benjamin:2021ygh} for more detail.} At $c=2$, a surprisingly intricate structure was shown with the overlap with Maass cusp forms \cite{Benjamin:2021ygh}:
\begin{equation}
(Z^{c=2}(\rho, \sigma), \nu_n) = 8\nu_n(\rho)\nu_n(\sigma),
\label{eq:miracle8s}
\end{equation}
where $\rho, \sigma$ are the complex structure and complexified K\"ahler moduli of the $c=2$ CFT. It would be interesting to repeat this analysis for higher $c$. In particular, it is curious that such a simple expression occurs on the RHS of (\ref{eq:miracle8s}).

The RHS of (\ref{eq:miracle8s}) can be thought of as a discrete eigenfunction of the Laplacian on the $c=2$ Narain moduli space. The overlap with the Maass cusp forms at higher $c$ must similarly be a discrete eigenfunction of the Laplacian on Narain moduli space. The only question is the normalization. At $c=2$ a particularly simple normalization occurs -- essentially just a factor of $8$ (note that the functions $\nu_n$ are not unit-normalized under the Petersson inner product; see \cite{Benjamin:2021ygh}). At higher $c$, one way to identify the normalization is to take a decompactification limit, where the discrete eigenfunctions have known normalization.

Similarly it would be nice to redo this analysis for higher genus partition functions (at any value of $c$).

\section*{Acknowledgments}
We are grateful to  Cyuan-Han Chang, Scott Collier, Jim Halverson and Ami Katz for useful discussions, and to Cameron Cogburn for collaboration in the early stages of the project.  We are also grateful to Boris Pioline for bringing to our attention the works \cite{dixon1991moduli,Kiritsis:1997em, Bossard:2016hgy} and to Eric Perlmutter for comments on a previous version of the paper. NB is supported by the U.S. Department of Energy, Office of Science, Office of High Energy Physics, under Award Number DE-SC0011632, and by the Sherman Fairchild Foundation. ALF is supported by the US Department of Energy Office of Science under Award Number DE-SC0015845, the Simons Collaboration on the Non-Perturbative Bootstrap, and a Sloan Foundation fellowship. ALF thanks the Aspen Center for Physics for hospitality while this work was completed.

\appendix

\section{Unfolding Narain CFT Partition Functions}
\label{app:orbits}

\subsection{Decomposition into Orbits}

In this appendix, we summarize the method in \cite{dixon1991moduli} (and explicitly generalized to $c>2$ in \cite{Kiritsis:1997em}) for summing over orbits of $SL(2,\mathbb{Z})$ of Narain lattice points; we will also describe in a bit more detail how to efficiently perform the sum over such orbits.  The idea is that there is a useful trick for evaluating the integral in (\ref{eq:Petersson}) in the case where $f$ (or $g$) can be written as a nontrivial sum over modular images of a different function $f_0(x,y)$.  That is, let $f$ be related to $f_0$ by
\begin{equation}
f(x,y) = \sum_{\gamma \in SL(2,\mathbb{Z}) /S_{f_0}} f_0(\gamma(x,y)),
\label{eq:UnfoldedF}
\end{equation}
where $S_{f_0}$ is the stabilizer of the function $f_0$ in $SL(2,\mathbb{Z})$, i.e.~$S_{f_0} \equiv \{ \gamma \in SL(2,\mathbb{Z}) | \gamma \cdot f_0 = f_0 \}$, $(\gamma \cdot f_0)(x,y) \equiv f_0(\gamma(x,y))$.   Then, the `unfolded' integral is
\begin{equation}
(f,g) = \int_{S_{f_0} \setminus H} \frac{dx dy}{y^2} f_0(x,y) g^*(x,y).
\end{equation}
This deceptively simple trick has surprisingly deep consequences. For one, the space $S_{f_0} \setminus H$ is often simper and easier to integrate over than $\mathcal{F}$.  As a result, the quantities appearing in the unfolded integral are often themselves more accessible to further analysis. The two  specific cases we will encounter are when $f_0$ is a function of $y$ only, so $S_{f_0} = \Gamma_\infty$ is the set of translations in $x$ by an integer; and when $f_0$ is not invariant under any nontrivial $SL(2,\mathbb{Z})$ transformation, so $S_{f_0}$ just contains the identity.  In the former case, $S_{f_0} \setminus H = \{ (x,y) | -\frac{1}{2} < x \le \frac{1}{2}, 0<y\}$ is the strip, and in the latter case $S_{f_0} \setminus H = H$ is the entire upper half-plane.

Our main interest is to apply this technique in the case where $f$ is the partition function of a Narain CFT.  We will use the fact that Narain CFT partition functions can be expressed as nontrivial sums of the form (\ref{eq:UnfoldedF}).  In general, the moduli space of Narain CFTs with central charge (equivalently, target space dimension) $c$ is parameterized by a symmetric $c \times c$ matrix $G_{ij}$ and an antisymmetric $c\times c$ matrix $B_{ij}$.  The spectrum of operator dimensions and spins at a fixed point in moduli space is given by
\begin{equation}
\Delta_{n,w} = \frac{M_{n,w}^2}{2} = \frac{p_a G^{ab} p_b + w^c G_{cd}w^d}{2} , \qquad \ell_{n,w} = n_a w^a, \qquad p_a \equiv n_a + B_{ac} w^c,
\end{equation}
where $n,w \in \mathbb{Z}^c$.  Their modular-invariant `primary' partition function is defined as
\begin{equation}
\hat{Z}^{(c)}(x,y) = y^{\frac{c}{2}} \sum_{n,w \in \mathbb{Z}^c} \exp\left( - \pi y M_{n,w}^2 + 2 \pi i x \ell_{n,w}\right).
\end{equation}
One can obtain a manifestly modular-invariant expression for $\hat{Z}^{(c)}$ through Poisson resummation:
\begin{equation}
\hat{Z}^{(c)}(x,y) = \sqrt{\det(G)} \sum_{m,n \in \mathbb{Z}} \exp\left( - \pi \frac{(m^i + n^i \tau) G_{ij} (m^j + n^j \bar{\tau})}{y} + 2 \pi i B_{ij}m^i n^j \right).
\label{eq:PoissResum}
\end{equation}
With this form, a modular transformation of the partition function $\hat{Z}^{(c)}(x,y) \rightarrow \hat{Z}^{(c)}(\gamma(x,y))$ can be seen to be equivalent to a rearrangement of terms in the sum, and so the full partition function is invariant.  Explicitly, each term in the sum is invariant if we transform $x,y$ according to $\tau \rightarrow \tau' = \gamma^{-1}(\tau) = \frac{d \tau-b}{a-c \tau}$ and simultaneously transform $n^i \rightarrow n'^i = a n^i+c m^i, m^i \rightarrow m'^i = b n^i + dm^i$.    In this way, the  ``lattice'' coordinates $n^i, m^i$ inherit an action under $SL(2,\mathbb{Z})$. 

To put  (\ref{eq:PoissResum}) into the form (\ref{eq:UnfoldedF}),  we can group the lattice coordinates into orbits  of $SL(2,\mathbb{Z})$, and break up the sum over all lattice coordinates into a product of two sums, one over the orbits and one within the orbits. Define
\begin{equation}
z(n,m;x,y) \equiv\sqrt{\det(G)} \exp\left( - \pi \frac{(m^i + n^i \tau) G_{ij} (m^j + n^j \bar{\tau})}{y} + 2 \pi i B_{ij}m^i n^j \right).
\label{eq:IndividualZ}
\end{equation}
Then, 
\begin{equation}
\hat{Z}^{(c)}(x,y) =  \sum_{\psi \textrm{ orbits}} z_\psi(x,y),  \quad z_\psi(x,y) \equiv \sum_{(n^i, m^i ) \in \psi} z(n^i, m^i; x,y).
\end{equation}
We will discuss the explicit form of the orbits $\psi$ shortly, but for now all that is important is that by definition, within a single orbit every lattice coordinate is related to every other one by a $SL(2,\mathbb{Z})$ transformation.  Therefore, we can rewrite each sum within an orbit as a sum over elements of $SL(2,\mathbb{Z})$.  To do this, for each orbit $\psi$, choose a representative element $(n_\psi^i, m_\psi^i)$, which is specific to that orbit.  Define $\Gamma_\psi$ to be the stabilizer in $SL(2,\mathbb{Z})$ of that element (i.e. $\gamma \in \Gamma_\psi \Leftrightarrow \gamma \cdot (n_\psi^i, m_\psi^i) = (n_\psi^i, m_\psi^i)$).  Now, for each element $(n^i, m^i)$ in the orbit $\psi$, there is a unique $\gamma \in SL(2, \mathbb{Z})/\Gamma_\psi$ such that $(n^i, m^i) = \gamma \cdot (n^i_\psi, m^i_\psi)$.  Consequently, 
\begin{equation}
z_\psi(x,y) = \sum_{\gamma \in SL(2,\mathbb{Z})/\Gamma_\psi} z(n_\psi^i, m_\psi^i, \gamma (x,y)).
\end{equation}
We have therefore written the partition function $\hat{Z}^{(c)}(x,y)$ as a sum over terms, each of which is individually of the form (\ref{eq:UnfoldedF}).  Given any other modular invariant function $g$, we now write the Petersson inner product $(\hat{Z}^{(c)},g)$ as a sum over the orbits of the lattice coordinates, and `unfold' each sum within an orbit:
\begin{equation}
(\hat{Z}^{(c)}, g) 
=\sum_{\psi \textrm{ orbits}} \int_{\mathbb{H}/\Gamma_\psi } \frac{dx dy}{y^2}  z(n_\psi^i, m_\psi^i, x,y) g^*(x,y).
\label{eq:ReducedNarainInnerProduct}
\end{equation}

\subsection{Explicit Description of Lattice Coordinate Orbits}

Next, we discuss the explicit form of the orbits of the lattice coordinates.  The stabilizers $\Gamma_\psi$ are independent of the choice of the representative elements $(n_\psi^i, m_\psi^i)$, but the explicit form of the integrand  $z(n_\psi^i, m_\psi, x,y)$ is not and so we will also want to choose representative elements that make performing the integration as simple as possible.  As an illustrative warm-up case, we will start with the cases $c=1$ and $c=2$, and then move on to general $c$.\footnote{The $c=1$ and $c=2$ analysis here is largely a repeat of that in appendix C of \cite{Benjamin:2021ygh}, though we will frame it in slightly more general terms.}

\subsubsection{$c=1$}

In the case of $c=1$, the lattice coordinates are just pairs $(n,m)$, and two coordinates are in the same orbit iff they have the same value for $\gcd(n,m)$.  To see this, note that if $\gcd(n,m) = m_\psi$, then $(n,m)$ can be obtained from an $SL(2,\mathbb{Z})$ transformation $n\rightarrow a n+c m, m \rightarrow bn+dm$ with $c=n/m_\psi, d= m/m_\psi$ acting on $(0,m_\psi)$.\footnote{There is always a possible choice of $a,b$ such that $ad-bc=1$, since $c$ and $d$ are coprime by virtue of $m_\psi = \gcd(n,m)$.}  Moreover, a convenient set of representatives for the orbits is $\{ (0,m_\psi) \}_{m_\psi \in \mathbb{Z}}$.  For any value of $m_\psi$, the stabilizer of the orbit is $\Gamma_\psi = \Gamma_\infty$,\footnote{Following convention, we define the subgroup $\Gamma_\infty \cong \left\{ \left( \begin{array}{cc} 1 & n \\ 0 & 1 \end{array} \right) \Big| n \in \mathbb{Z} \right\}$.} except for $m_\psi=0$ where $\Gamma_\psi = SL(2,\mathbb{Z})$.  Therefore, for $c=1$, (\ref{eq:ReducedNarainInnerProduct}) reduces to
\begin{equation}
(\hat{Z}^{(c=1)},g) = \int_{\mathbb{F}} \frac{dxdy}{y^2} z^{(c=1)}(0,0,x,y)g^*(x,y) + \sum_m' \int_{\mathbb{H}/\Gamma_\infty} \frac{dx dy}{y^2} z^{(c=1)}(0,m,x,y) g^*(x,y),
\end{equation}
where the prime on the sum indicates that the $m=0$ term is omitted. In the case $c=1$, it is common to define $R^2 = G_{11}$, so
\begin{equation}
z^{(c=1)}(0,m;x,y) = R \exp\left(-\pi \frac{m^2 R^2}{y}\right).
\end{equation}
Then, we can simplify (\ref{eq:ReducedNarainInnerProduct})  even further, to
\begin{equation}
(\hat{Z}^{(c=1)},g) = R \int_{\mathbb{F}} \frac{dx dy}{y^2} g^*(x,y) + R \sum'_m \int_{-\frac{1}{2}}^{\frac{1}{2}} dx \int_0^\infty \frac{dy}{y^2} e^{- \pi \frac{m^2 R^2}{y}} g^*(x,y).
\label{eq:C1Unfolded}
\end{equation}
Although this may not look like much of an improvement, we will see that this form for $(\hat{Z}^{(c=1)},g)$ is much more amenable to explicit evaluation than the one we started with.  The second term on the RHS above is simpler because the $x$ and $y$ integration regions are factorized, and the first is simpler because it contains only a single term from $\hat{Z}^{(c=1)}$ rather than a sum over lattice coordinates or states.  

\subsubsection{$c=2$}

Next we consider the case $c=2$.  There are now two sets of pairs, $(n^1, m^1)$ and $(n^2, m^2)$, and it will prove convenient to group them into a two-by-two matrix $N_{12}$ with determinant $D_{12}$:
\begin{equation}
N_{12} =  \left( \begin{array}{cc} n^1 & n^2 \\ m^1 & m^2 \end{array} \right) , \qquad D_{12}  \equiv\det N_{12}= m^1 n^2 - m^2 n^1.
\label{eq:D12}
\end{equation}
This determinant is invariant under $SL(2,\mathbb{Z})$ transformations of the $n^i, m^i$s.  Moreover, the stabilizer $\Gamma_\psi$ is trivial for any orbit with nonzero $D_{12}$.\footnote{The action of an $SL(2,\mathbb{Z})$ element $\gamma$ acting on the lattice coordinates can be written as left multiplication of $\gamma^t$ times $N_{12}$ in (\ref{eq:D12}).  If $D_{12}\ne 0$, then $N_{12}$ is invertible, so $\gamma^t \cdot N_{12}= N_{12}$ implies that $\gamma$ is the identity element.}  Treating the $(n^1, m^1)$ coordinates as in the $c=1$ case, we can always find an $SL(2,\mathbb{Z})$ transformation that sets $n_1=0$.  If $D_{12} \ne 0$, then after performing this transformation, $n^2$ cannot vanish, and by an $SL(2,\mathbb{Z})$ transformations of the form $n^i \rightarrow n^i, m^i \rightarrow m^i + b n^i$, we can always shift $m_2$ to be between 0 and $n^2-1$. Thus for $D_{12}\ne 0$ we can always bring $N_{12}$ into the form
\begin{equation}
N_{12} \cong \left( \begin{array}{cc} 0 & n_\psi^2 \\ m_\psi^1 & m_\psi^2 \end{array} \right), \qquad \textrm{ with } n_\psi^2 m_\psi^1 = -D_{12}, \quad n_\psi^2>0, \textrm{ and } 0 \le m_\psi^2 \le n_\psi^2 -1,
\label{eq:c2NonZeroRepresentatives}
\end{equation}
 and that no two distinct $N_{12}$s of this form are related by a nontrivial $SL(2,\mathbb{Z})$ transformation.  Therefore, (\ref{eq:c2NonZeroRepresentatives}) can be taken as a complete set of representative elements for the orbits with $D_{12} \ne 0$.  
 
 On the other hand, if $D_{12}=0$, then for any $N_{12}$, let $u^i \equiv \gcd(n^i, m^i)$, and $(c^i, d^i) = (n^i/u^i, m^i/u^i)$. By construction, $\gcd(c^i, d^i)=1$, so $D_{12}=0$ implies $c^1=c^2\equiv c$ and  $d^1 = d^2\equiv d$.  This means that, $(n^1, m^1)$ and $(n^2, m^2)$ are obtained from $(0, u^1)$ and $(0,u^2)$, respectively, by the same $SL(2,\mathbb{Z})$ transformation.  Therefore, a complete set of representatives of the orbits with $D_{12}=0$ are
 \begin{equation}
 N_{12} \cong \left( \begin{array}{cc} 0 & 0 \\ m_\psi^1 & m_\psi^2 \end{array} \right), \qquad m_\psi^1, m_\psi^2 \in \mathbb{Z}.
 \end{equation}

For $c=2$, it is convenient to reparameterize the moduli space in terms of two complex coordinates $\sigma= \sigma_1+i \sigma_2$ and $\rho=\rho_1+i \rho_2$, defined as
\begin{equation}
\rho \equiv B_{12}+ i \sqrt{\det G}, \quad \sigma \equiv \frac{G_{12}}{G_{11}} + i \frac{\sqrt{\det G}}{G_{11}}  \Leftrightarrow G = \frac{\rho_2}{\sigma_2} \left( \begin{array}{cc} 1 & \sigma_1 \\ \sigma_1 & |\sigma|^2 \end{array} \right), \quad B= \left( \begin{array}{cc} 0 & \rho_1 \\ -\rho_1 & 0 \end{array} \right).
\label{eq:C2Modulus}
\end{equation}
In these variables, the basic elements $z^{(c=2)}$ from (\ref{eq:IndividualZ}) take the simple form
\begin{equation}
\frac{z^{(c=2)}(n^i, m^i; x,y)}{\rho_2} =  \exp\left(  \frac{i \pi }{2} \left( \frac{\rho |m^2 + m^2 \bar{\sigma} + \tau (n^1 + n^2 \bar{\sigma})|^2 }{y \sigma_2} - \frac{\bar{\rho} |m^1 + m^2 \sigma + \tau (n^1 + n^2 \sigma)|^2}{y \sigma_2} \right) \right).
\end{equation}
Finally, substituting our choice of representative elements for each orbit, our formula (\ref{eq:ReducedNarainInnerProduct}) applied to $c=2$ reduces to 
\begin{equation}
\begin{aligned}
\frac{(\hat{Z}^{(c=2)},g)}{\rho_2} &= \int_{\mathbb{F}} \frac{dxdy}{y^2}g^*(x,y) + \sum_{m_1, m_2}' \int_{\mathbb{H}/\Gamma_\infty} \frac{dx dy}{y^2} e^{- \pi \frac{\rho_2}{\sigma_2} \frac{|m_1 + m_2 \sigma|^2}{y} } g^*(x,y) \\
 &+\sum_{D}' \sum_{n_2 | D} \sum_{m_2=0}^{n_2-1} \int_{\mathbb{H}} \frac{dx dy}{y^2}e^{ \frac{i \pi}{2} \left( \frac{\rho | \frac{D}{n_2} + \bar{\sigma} (m_2+n_2 \tau)|^2}{y \sigma_2}  -  \frac{\bar{\rho} |\frac{D}{n_2}+ \sigma (m_2+n_2 \tau)|^2}{y \sigma_2}  \right) }  g^*(x,y).
\end{aligned}
\end{equation}

\subsubsection{$c \ge 3$}

We finally turn to the general case with $c\ge 3$.  We again define a matrix $N_{ij}$ containing all the lattice coordinates:
\begin{equation}
N_{ij} = \left( \begin{array}{cccc} n^1 & n^2 & \dots & n^c \\ m^1 & m^2 & \dots & m^c \end{array} \right).
\label{eq:LatticeCoords}
\end{equation}
Now there are $\frac{c(c-1)}{2}$ different determinants $D_{ij} \equiv \det N_{ij}$, which are all invariant under $SL(2,\mathbb{Z})$ transformations.  We will organize all the allowed orbits first by the values of $D_{ij}$.

If all $D_{ij}$ determinants vanish, then there is an $SL(2,\mathbb{Z})$ transformation that sets all the $n^i$s to zero, by the same argument we used for $c=2$ when $D_{12}=0$. We can therefore choose a representative set of elements for this case to be 
\begin{equation}
N_{ij} = \left( \begin{array}{cccc} 0 & 0 & \dots & 0 \\ m_\psi^1 & m_\psi^2 & \dots & m_\psi^c \end{array} \right), \qquad m_\psi^i \in \mathbb{Z}^c .
\label{eq:AllDsVanish}
\end{equation}
The stabilizer $\Gamma_\psi$ of the orbit is $\Gamma_\infty$,  unless all $m_\psi^i =0$ in which case $\Gamma_\psi = SL(2,\mathbb{Z})$.  

As long as at least one of the $D_{ij}$ does not vanish, we can focus on one of the $N_{ij}$s with nonzero determinant.  For concreteness, let us say $D_{12} \ne 0$, and transform $N_{12}$  into the form (\ref{eq:c2NonZeroRepresentatives}).  Now, note that once $N_{12}$ is fixed, knowledge of the $D_{ij}$s is sufficient to fix all the remaining $m^i, n^i$s:
\begin{equation}
n_\psi^i = \frac{D_{1i} n_\psi^2}{D_{12}} = - \frac{D_{1i}}{m_\psi^1}, \quad m_\psi^i = \frac{m_\psi^2 D_{1i} - m_\psi^1 D_{2i}}{D_{12}}.
\end{equation}
Of course, it may be that case that the $m_\psi^i, n_\psi^i$s determined this way are not integers, which means that particular case of $N_{12}$ is not consistent with the remaining $D_{ij}$s and should not be included in the sum over orbits. In fact, most choices of the set of determinants $D_{ij}$ will not be consistent with {\it any} choice of lattice coordinates.  In particular, note that $n_\psi^i = -D_{1i}/m_\psi^1$ implies that all the determinants $\{ D_{1j}\}_{j=1, \dots, c}$ must have a common divisor.  Moreover, by running the same argument with any other choice of a 2-by-2 submatrix in place of $N_{12}$, we see that the determinants $\{ D_{ij} \}_{j=1, \dots, c}$ must have a common divisor for any choice of $i$. As a practical matter, for a fixed set of determinants $D_{ij}$, we can choose the index $a$ to minimize $\gcd \left( \{ D_{aj} \}_{j=1, \dots, c} \right)$, and then we can choose the index $b$ to minimize $D_{ab}$, and replace $N_{12}$ above with $N_{ab}$.  

We can therefore write the Petersson inner product with a $c\ge 3$ partition function as
\begin{equation}
\begin{aligned}
\frac{(\hat{Z}^{(c\ge 3)}(x,y),g)}{\sqrt{\det G}}& = \int_{\mathbb{F}} \frac{dx dy}{y^2} g^*(x,y) + \sum'_{m_\psi^i \in \mathbb{Z}^c} \int_{\mathbb{H}/\Gamma_\infty} \frac{dx dy}{y^2} z^{(c\ge 3)} (0, m_\psi^i; x,y) g^*(x,y) \nn\\
& + \sum'_{\textrm{ consistent }D_{ij}}\sum_{n_\psi^i(D), m_\psi^i(D)} \int_{\mathbb{H}} \frac{dx dy}{ y^2} z^{(c>3)}(n_\psi^i(D), m_\psi^i(D); x,y) g^*(x,y).
\end{aligned}
\end{equation}
The final term in the sum above deserves some clarification.  The prime on the sum indicates that we do not include the term with all $D_{ij}$ vanishing.   Based on our discussion above, there are also many choices of $D_{ij}$ which are not consistent, and we do not include those either.  Moreover, we can readily tell if a given choice of $D_{ij}$ is consistent, and simultaneously pick a representative set of lattice coordinates $(n_\psi^i, m_\psi^i)$, as follows.  Among all sets of $\{D_{aj}\}_{j=1, \dots, c}$, choose the value of $a$ that minimizes $\gcd\left( \{D_{aj}\}_{j=1, \dots, c} \right)$, and then choose the value of $b$ that minimizes $|D_{ab}|$ among the nonzero values of $D_{ab}$.\footnote{If there are two values of $a$ that give the same $\gcd$, we can pick the one that gives the smaller value of $|D_{ab}|$; if that still leaves multiple values of $a$ or $b$ to choose from, we can arbitrarily take the one with the smallest value of $a$, followed by the smallest value of $b$.} Then, the sum over $(n_\psi^i(D), m_\psi^i(D))$ is a sum over all $m_\psi^a$ and $m_\psi^b$ such that
\begin{equation}
 m_\psi^a \textrm{ divides } D_{aj} \forall j, \qquad n_\psi^b = -\frac{D_{ab}}{m_\psi^a} >0, \qquad m_\psi^b = 0, \dots, n_\psi^b-1.
 \end{equation}
Then, all other lattice coordinates are given by
\begin{equation}
n_\psi^i = \frac{D_{ai} n_\psi^b}{D_{ab}} = - \frac{D_{ai}}{m_\psi^a}, \quad m_\psi^i = \frac{m_\psi^b D_{ai} - m_\psi^a D_{bi}}{D_{ab}}.
\end{equation}
Finally, if any of the resulting $m_\psi^i$ are not integers, then that set of $\{D_{ij}\}$ is not consistent and is not included in the sum; otherwise, it is.

\section{Recursion Formula for Epstein Zeta Function}
\label{app:Terras}

In \cite{terras1973bessel},  
Terras derived the formula (\ref{eq:TerrasFormula}) for $E_n(G,s)$,
\begin{equation}
E_n(G,s) \equiv \frac{1}{2} \int_0^\infty \frac{dy}{y} y^s \sum_{a \in \mathbb{Z}^n}' e^{- \pi y a^i G_{ij} a^j}.
\label{eq:EnDef2}
\end{equation}
 Here we briefly summarize her derivation and explain the notation.   The original metric is first decomposed as follows:
\begin{equation}\begin{aligned}
G = \left( \begin{array}{cc} G_1 & G_{12} \\ G_{12}^t & G_{2} \end{array} \right) \equiv  \left( \begin{array}{cc} 1 & Q \\ 0 & 1 \end{array} \right) \left( \begin{array}{cc} T_1 & 0 \\ 0 & G_{2} \end{array} \right)   \left( \begin{array}{cc} 1 & Q \\ 0 & 1 \end{array} \right)^t
\end{aligned}\end{equation}
where $G_1, G_2$ are $n_1 \times n_1$ and $n_2 \times n_2$ matrices, respectively, with $n = n_1 + n_2$. 
Clearly,
\begin{equation}\begin{aligned}
Q \equiv G_{12} G_{2}^{-1}, \qquad T_1 \equiv G_1 - Q G_{2} Q^t = G_1 - G_{12}G_2^{-1} G_{12}^t
\end{aligned}\end{equation}
We also introduce notation for the inner product of a vector $a$ with respect to a metric $G$:
\begin{equation}\begin{aligned}
G[a] \equiv a \cdot G \cdot a^t .
\end{aligned}\end{equation}
Note that
\begin{equation}\begin{aligned}
G[a] = T_1[a_1] + G_2[a_2 + Q a_1], \qquad \vec{a} \equiv (\vec{a}_1, \vec{a}_2)
\end{aligned}\end{equation}
Consequently, in the sum over $\vec{a}$s in $E_n(G,s)$, it is useful to separate out the terms with $\vec{a}_1=\{0, \dots, 0\}$:
\begin{equation}
\begin{aligned}
E_n(G,s) &= \frac{1}{2} \int_0^\infty \frac{dy}{y} y^s\left[ \sum_{a_2 \in \mathbb{Z}^{n_2} - 0} e^{ - \pi y ( G_2[a_2 ])} + \sum_{a_1 \in \mathbb{Z}^{n_1} - 0 \atop a_2 \in \mathbb{Z}^{n_2} } e^{ - \pi y ( T_1[a_1] + G_2[a_2 + Q a_1])} \right] \\
& = E_{n_2}(G_2, s) +  \frac{1}{2} \int_0^\infty \frac{dy}{y} y^s \sum_{a_1 \in \mathbb{Z}^{n_1} - 0 \atop a_2 \in \mathbb{Z}^{n_2} } e^{ - \pi y ( T_1[a_1] + G_2[a_2 + Q a_1])} 
\end{aligned}
\end{equation}
Next, one uses the Poisson resummation formula for the sum over $\vec{a}_2$:
\begin{equation}
\sum_{ a_2 \in \mathbb{Z}^{n_2} } e^{ - \pi y  G_2[a_2 + Q a_1]}  = \frac{1}{\sqrt{\det(G_2)}}y^{-\frac{n_2}{2}}  \sum_{ b_2 \in \mathbb{Z}^{n_2} } e^{ - \pi y^{-1}  G_2^{-1}[b_2]- 2\pi i b_2^i Q_{ij} a^j_1} .
\end{equation}
Finally, one separates out the contribution  from $\vec{b}_2=\{0, \dots, 0\}$, which is proportional to $E_{n_1}(T, s-\frac{n_2}{2})$, and the contribution from terms with $\vec{b}_2 \ne \{0, \dots, 0 \}$.  In the latter case, the integral over $y$ can be performed term-by-term in the sum over $\vec{a}_1$ and $\vec{b}_2$, with the result shown in (\ref{eq:TerrasFormula}).

To evaluate $E_n(G,s)$ recursively with $n=3$, we take $n_1=1$ and $n_2=2$ (or equivalently, $n_1=2, n_2=1$).   In the special case where $n_1=1$, we have
\begin{equation}\begin{aligned}
T_1 = G_{11} - G_{12} G_2^{-1} G_{12}^{t} \cong g_\perp, \qquad G_2 \cong h, \qquad n_2 = n-1 \cong c-1
\end{aligned}\end{equation}
and
\begin{equation}\begin{aligned}
E_{n_1} (T_1, s) =\frac{\pi^{-s}\Gamma(s)}{2} \sum_{m\ne 0} (T_1 m^2)^{-s} = T_1^{-s} \Lambda(s)
\end{aligned}\end{equation}
so for $n_1=1$, we have
\begin{equation}\begin{aligned}
E_n(G,s)  =&E_{n-1}(G_2, s) +  \frac{\Lambda(s- \frac{n-1}{2}) }{\sqrt{\det G_2}} T_1^{-s+\frac{n-1}{2}} \\
        & +  \frac{1}{\sqrt{\det G_2}}  \sum_{a_1 \in \mathbb{Z} - 0 \atop b_2 \in \mathbb{Z}^{n-1}-0 } \left( \frac{G_2^{-1}[b_2]}{T_1 a_1^2} \right)^{\frac{2s-(n-1)}{4}}  e^{- 2 \pi i b_2 Q a_1} K_{s - \frac{n-1}{2}} (2\pi \sqrt{ T_1 a_1^2 G_2^{-1}[b_2]}) \nn
\end{aligned}\end{equation}
Then for $n=3$, we can evaluate this expression explicitly using the fact that the  case $n=2$ is proportional to the usual real analytic Eisenstein series $E_s(\tau)$:
\begin{equation}
E_2(G,s) =\Lambda(s) (\det(G))^{-s/2} E_s(\sigma), \qquad \sigma \equiv \frac{G_{12}}{G_{11}} + i \frac{\sqrt{ \det G}}{G_{11}}.
\end{equation}

\bibliographystyle{JHEP}
\bibliography{harmonicAnalysisbib}
\end{document}